\documentclass[journal]{IEEEtran}

\usepackage{cite}
\usepackage{xcolor}

\usepackage[absolute]{textpos}

\ifCLASSINFOpdf
\else
   \usepackage[dvips]{graphicx}
   \graphicspath{{../gfx/}}
\fi

   \usepackage{psfrag}
    \usepackage{url}

\usepackage{amsmath}

\newcommand{\copyrightstatement}{
    \begin{textblock}{15}(0.5,0.7)    
         \noindent
         \centering
         \textblockcolour{white}
         \footnotesize
         \copyright 2022 IEEE. Personal use of this material is permitted. Permission from IEEE must be obtained for all other uses, in any current or future media, including reprinting/republishing this material for advertising or promotional purposes, creating new collective works, for resale or redistribution to servers or lists, or reuse of any copyrighted component of this work in other works
    \end{textblock}}

\begin{document}
\title{Sweet Streams are Made of This: \\ The System Engineer's View on Energy Efficiency in Video Communications }
\copyrightstatement

\author{Christian~Herglotz,~\IEEEmembership{Member,~IEEE,}
        Matthias Kr\"anzler,~\IEEEmembership{Member,~IEEE,} \\
        Robert Schober,~\IEEEmembership{Fellow,~IEEE,}
        and~Andr\'e Kaup,~\IEEEmembership{Fellow,~IEEE}
\thanks{C. Herglotz, M. Kr\"anzler, and A. Kaup are with the Chair of Multimedia Communications and Signal Processing, R.  Schober is with the Institute for Digital Communications, Friedrich-Alexander University Erlangen-N\"urnberg (FAU),
Erlangen, Germany e-mail: \{christian.herglotz, matthias.kraenzler, andre.kaup, robert.schober\} @fau.de.}
\thanks{Manuscript received April 30, 2021; revised December 20, 2021; accepted September 1, 2022.}
\thanks{This work was partially funded by the Deutsche Forschungsgemeinschaft (DFG, German Research Foundation), project number 447638564.}}

\markboth{Circuits and Systems Magazine, preprint}{}%

\maketitle

\begin{abstract}
In recent years, the global use of online video services has increased rapidly. Today, a manifold of applications, such as video streaming, video conferencing, live broadcasting, and social networks, make use of this technology. A recent study found that the development and the success of these services had as a consequence that, nowadays, more than $\mathbf{1\%}$ of the global greenhouse-gas emissions are related to online video, with growth rates close to $\mathbf{10\%}$ per year. This article reviews the latest findings concerning energy consumption of online video from the system engineer's perspective, where the system engineer is the designer and operator of a typical online video service. We discuss all relevant energy sinks, highlight dependencies with quality-of-service variables as well as video properties, {\color[rgb]{0,0,0} review energy consumption models for different devices from the literature, and aggregate these existing models into a global} model for the overall energy consumption of a generic online video service. Analyzing this model and its implications, we find that end-user devices and video encoding have the largest potential for energy savings. Finally, we provide an overview of recent advances in energy efficiency improvement for video streaming and propose future research directions for energy-efficient video streaming services. 

\end{abstract}

\IEEEpeerreviewmaketitle
 

\section{Introduction}
\IEEEPARstart{W}{ith} the advent of portable devices and powerful video compression techniques, on-demand video streaming and communication services have become an integral part of the daily lives of billions of users all over the world within the last decade \cite{Malmodin20}. Moreover, due to substantial advances in digital data transmission technology, the demand for classical video services such as analog television broadcast is declining \cite{statistaTV}, as the bandwidth of networks is sufficiently high to send videos separately to individual users, who are getting used to being able to watch the content on-demand anytime and anywhere. 

As a downside to this development, it was found that the energy consumption, which is directly related to the production of greenhouse gases (GHG) causing climate change \cite{ShiftFull19}, has simultaneously increased dramatically. A recent study from the year 2019 claims that for the year 2017, more than $1\%$ of the global GHG emissions can be attributed to online video services \cite{ShiftFull19}, which was equal to a quarter of the emissions caused by global aviation. These emissions can be attributed to the production of related devices, which makes up $45\%$ of the online video-related emissions, and the actual use of online video services, which comprises the remaining $55\%$ and which we focus on in this work. According to 
\cite{ShiftFull19}, 
the demand for online video services, including the overall data rate required for video delivery, will further increase over the next years, which is consistent with the findings of \cite{cisco19,cisco20, Sandvine20}.

Already for the year 2019, a substantial and strong growth in transmitted video data of more than $8\%$ was expected \cite{ShiftFull19}. With the appearance of the Corona virus, {\color[rgb]{0,0,0} the demand for online communications grew unexpectedly \cite{Delgado21} such that}, in March 2020 alone, the Internet traffic at Germany's biggest Internet exchange point (IXP) increased by $10\%$. At the same time, traffic related to online video conferencing increased by $50\%$ \cite{DECIX}, showing that the actual growth rate {\color[rgb]{0,0,0}of online video services} will likely be much higher than previously predicted.

In the literature, one can find a plethora of studies investigating the overall impact of online video technologies as well as other information and communication technologies (ICT) on the global energy consumption and GHG emissions \cite{Malmodin20,ShiftFull19,Hintemann16,Belkhir18,Yan19,Shehabi16}.
Furthermore, some studies attempt to break down the overall energy consumption to the end-user level, where the target is to provide end users with information on the GHG emissions caused by the streaming of a single video \cite{Malmodin20,Malmodin16,ShiftFull19,umweltbundesamt20,Suski20}. In contrast, in this work, we target the overall energy consumption related to online video services, where we take the viewpoint of a system engineer constructing and maintaining a certain online video service. 

\begin{figure*}[h!t]
\centering
\psfrag{I}[c][c]{IXP}
\psfrag{T}[l][l]{\textbf{Transmission Network }($E_\mathrm{NW}$)}
\psfrag{4}[r][r]{(4G / 5G)}
\psfrag{M}[r][r]{Mobile Network}
\psfrag{D}[c][c]{\textbf{Terminals}}
\psfrag{E}[c][c]{\textbf{End-user} }
\psfrag{K}[c][c]{($E_\mathrm{UT}$)}
\psfrag{P}[c][c]{\textbf{Video}}
\psfrag{Z}[c][c]{\textbf{Providers} ($E_\mathrm{VP}$)}
\psfrag{G}[l][l]{Optical}
\psfrag{F}[l][l]{fiber}
\psfrag{C}[l][l]{Copper}
\psfrag{W}[l][l]{Wi-Fi}
\includegraphics[width=.85\textwidth]{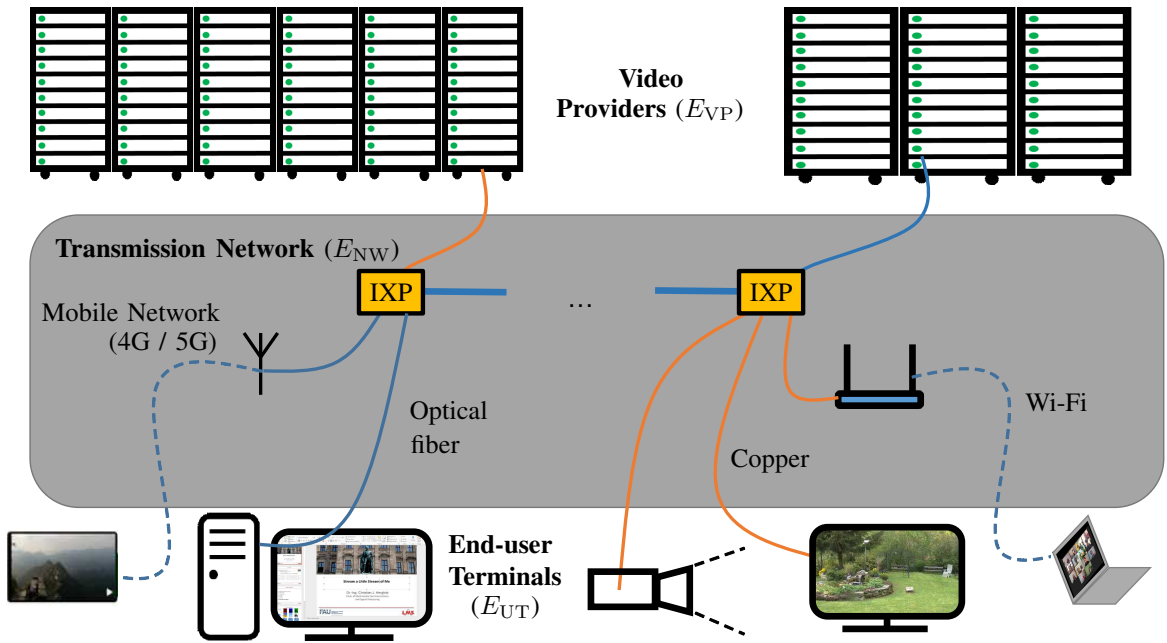} 
\caption{Active components in online video applications comprise data centers and the infrastructure of the online video providers (VP), transmission networks (NW), and end-user terminals (UT). The overall energy consumption of all components is denoted by the variable $\boldsymbol{E}$. The video providers store video data and respond to user requests using servers in data centers. The transmission networks include different transmission technologies using components such as Internet exchange points (IXPs), base stations, switches, and (home) routers. The interconnecting lines represent wireless (dashed lines) and wired (solid lines) connections. The end-user terminals comprise playback and recording devices such as smartphones, tablet PCs, desktop PCs, TV sets, and cameras. All components contribute to the overall energy consumption caused by online video applications.  }
\label{fig:overview}
\end{figure*}

As a consequence, we provide a generalized overview on the energy consumption of all devices and systems that perform online video tasks, which include data centers storing and providing videos, data transmission networks comprised of network nodes, and end-user devices such as TVs, smartphones or tablet PCs \cite{Malmodin20,Suski20}. We intend to cover all major applications such as (on-demand) streaming, video conferencing, social networks, broadcasting, and surveillance. 
To this end, we map all applications to a general system model that comprises the most common online video services. This model is depicted in Fig.~\ref{fig:overview}. The system model contains three main components: The first component comprises terminals, which are end-user devices operated by users who request video streams \cite{Malmodin20, Herglotz20}. The second component includes all devices enabling the network connection for video data transmission. The networks transport the data using different kinds of data carriers such as Wi-Fi, 4G, and 5G for wireless transmission as well as copper or optical fiber for wired transmission. Typical devices in the transmission network are IXPs, switches, and routers. 
Finally, the third component mainly consists of data centers (DCs), where video data is transcoded, stored, and sent to the end users \cite{Bianco16,Wu20}. 

In this respect, we adopt the system engineer's perspective of an online video service. This means that we consider the complete infrastructure, i.e., all relevant devices mentioned above. 
The proposed system model allows to estimate the overall power and (yearly) energy consumption caused by a single online video service, e.g., a provider of video conferencing systems or an on-demand streaming platform. {\color[rgb]{0,0,0} We note that since we consider a large, distributed system consisting of thousands of independent devices located all over the world, validation measurements are extremely complex and costly. Hence, in this initial work, we rely on models for subsystems which have already been validated in the literature, such that the obtained energy consumption estimates are sufficiently accurate.

The proposed global model enables us} to reveal the main reasons for the high energy consumption of online video services, to point out new research directions, and to reveal opportunities for improving the energy efficiency in the future. {\color[rgb]{0,0,0} As such, the proposed global energy consumption model can be used as a baseline for future research on online video services, where researchers can determine the benefits of their work with respect to the global energy consumption. }

The remainder of this article is organized as follows. In Section~\ref{sec:global}, a thorough literature review will show the link between GHG emissions and the energy consumption of the devices required for online video service provisioning. Furthermore, we give an overview of the current research on the energy efficiency of data centers, transmission networks, and end-user devices.  
Afterwards, Section~\ref{sec:local} dives into a detailed system-level analysis to construct an energy consumption model based on the general system model shown in Fig.~\ref{fig:overview}. The resulting model describes and covers all major energy sinks that are present in video streaming. Furthermore, parameter values will be derived from the literature facilitating concrete energy estimates.
Finally, Section~\ref{sec:energyEfficiency} considers use cases for the proposed model, where we identify
 the most energy-intensive components of different online video services. Furthermore, we shortly describe state-of-the-art techniques and approaches to improve the energy efficiency of online video technology and present upcoming trends as well as potential future research directions targeting an increased energy efficiency. Finally, Section~\ref{sec:concl} concludes this paper.

\section{Global Energy Consumption in Online Video}
\label{sec:global}
In this section, we first discuss how the energy consumption of online video services ultimately leads to climate change in Subsection~\ref{secsec:GHG}. Afterwards, in Subsections~\ref{secsec:globDC} to \ref{secsec:globUT}, we provide an overview of the current literature on the energy consumption and energy modeling for provider-side data centers, transmission networks, and end-user devices as shown in Fig.~\ref{fig:overview}. 

\subsection{Energy, Greenhouse Gases, and Climate Change}
\label{secsec:GHG}

In science, it is well accepted that GHG emissions cause climate change, which is supported by many scientific studies, simulations, and surveys \cite{Boer00, Crippa19, Althor16}. In this respect, carbon dioxide (CO$_2$) is generally considered to be the most important factor, however, also other gases such as methane and nitrous oxide need to be considered \cite{Crippa19}. To simplify considerations, a unified metric was proposed, which translates all GHG emissions caused by any gas to the so-called CO$_2$-equivalents, which, in the following, will be referred to by the unit $[\mathrm{CO_2E}]$ \cite{Eggleston06}.

During the use of online video, the resulting GHG emissions are caused by the power consumption of all involved devices. In the literature, the conversion from power to emissions is usually done using the CO$_2$ or carbon intensity of energy production, which is the amount of CO$_2$ emissions in grams per energy unit $\left[ \frac{\mathrm{gCO_2E}}{\mathrm{kWh}}\right]$, where the energy is the time integral of the power \cite{ourworldindata}. This unit mainly depends on the type of power plant used to generate electricity. For example, modern wind parks have a very small CO$_2$-intensity that is close to $0\, \frac{\mathrm{gCO_2E}}{\mathrm{kWh}}$. In contrast, coal-fired power plants have a very high intensity of more than $350\, \frac{\mathrm{gCO_2E}}{\mathrm{kWh}}$ \cite{Quaschning19}. 
{\color[rgb]{0,0,0}Average carbon intensities for various countries }can be found online \cite{ourworldindata,electricitymap}. 

\subsection{Data Centers}
\label{secsec:globDC}

There is a large body of literature on the energy consumption and the processing efficiency of data centers (DCs)
\cite{Shehabi16,Borderstep20,Prakash14,Wang12,Uzaman19,Fiandrino15,Greenpeace17}. 
Most of this research targets general DCs that provide different kinds of services such as cloud services, storage, processing power for complex calculations, block chains, and so on.  

Various studies have shed light on the overall power consumption of nation-wide DCs, where detailed data is available especially for the US \cite{Shehabi16} and Europe \cite{Borderstep20, Prakash14}. The main findings from these studies are that the overall power consumption of DCs is constantly increasing over the years and that throughout the last decade, the power consumption of all DCs had a global share of the total energy consumption of $1\%$ to $2\%$. In these studies, the focus lies on the power consumption of the main hardware components, i.e., servers, storage units, and networking devices. Additionally, the power consumption related to infrastructure such as lighting and climatization is also considered \cite{Prakash14, Wang12}. 

To determine the energy efficiency of these DCs, the most common approach is to use the so-called power-usage efficiency (PUE) or its inverse, the data center infrastructure efficiency (DCE) \cite{Uzaman19, Fiandrino15, Wang12}. These figures of merit describe the relation between the power consumption of all computationally active components of a DC and the total power consumption including management and climatization. DCs are energy efficient if their PUE is close to one, which means that almost all the power is consumed by active components. The authors in \cite{Uzaman19} report that today, the DCs of the big tech companies have PUEs between $1.08$ and $1.45$. As the PUE does not provide information on the true power consumption and the related GHG emissions, further studies investigated the actual GHG emissions in detail \cite{Greenpeace17}.

An exhaustive overview of energy and power modeling approaches for DCs is given in \cite{Dayarathna16}. In this survey, power models are listed for many different components such as processors, storage, and software. Also, the total power consumption of an entire DC is discussed. For our proposed generic power consumption model for online video applications, in Section~\ref{sec:local}, we will select appropriate models from the aforementioned papers \cite{Shehabi16,Borderstep20,Prakash14,Wang12,Uzaman19,Fiandrino15,Greenpeace17,Dayarathna16}
 while taking into account the PUE.

\subsection{Transmission Networks}
\label{secsec:globNW}

Since the beginning of the '90s, the energy consumption of communication networks has been investigated in detail \cite{Malmodin20, Malmodin16}. Power and energy consumption data are available for many countries, operators, and technologies {\color[rgb]{0,0,0}\cite{Malmodin20, Coroama13, Morley18,Mammela17}}. In general, we may distinguish between the core network, in which data is transmitted globally between Internet exchange nodes, and access networks located close to the end user. For the former, it is common to consider the number of hops, where one hop represents the transmission from one Internet node to the next. Examples of such exchange nodes can be commercial Internet exchange points (CIX) or regional routers \cite{Coroama13}. Common to nodes in the core network is that they can potentially serve any Internet user worldwide. 

Access networks, which are located close to the end user, comprise fixed broadband access networks and mobile access networks \cite{Malmodin20}. Both provide Internet access to end users using either fixed lines (copper or optical fiber) or mobile networks (2G/3G/4G/5G). To obtain power consumption values for these networks, it is common to consider the overall power consumption of each network node involved in the transmission \cite{Malmodin20, Coroama13, Morley18, Alsharif19}. 

To quantify the energy efficiency of data transmission, a common method is to determine the energy intensity, i.e., the energy consumption per transmitted bit in $\left[\frac{\mathrm{J}}{\mathrm{bit}}\right]$ or the power consumption per throughput in $\left[\frac{\mathrm{W}}{\mathrm{bps}}\right]$ \cite{ShiftFull19,Coroama13, Gyli11}, where the unit bps represents ``bits per second''. Here, the power and the energy comprise the electromagnetic radiation power plus the additional power needed to operate the devices. Over the last decades, the energy and power intensities have decreased by several orders of magnitude 
\cite{Coroama13}. For example, the earliest estimates suggested a power  intensity of more than $3\,\frac{\mathrm{kW}}{\mathrm{Mbps}}$ before the year 2000, but only $36\,\frac{\mathrm{W}}{\mathrm{Mbps}}$ in 2016 \cite{Malmodin20, Preist19}. 

Recently, it was proposed to extend this simplified view, which just sets the overall energy consumption in relation to the transmitted data, by also taking into account other factors such as the number of users, subscribers, and lines \cite{Malmodin20}. It was argued that the traditional approach neglected that most of the power consumption is independent of the actual amount of transmitted data. This power is referred to as a constant offset power in the following, which is always consumed independently of the usage of the network. For example, a base station for a mobile network always consumes a certain offset power, even if no subscriber is logged in. Although additional power is consumed during data transmission, this additional power is often much smaller than the constant offset power \cite{Malmodin20}. Hence, it is argued that the constant part of the power should be associated with the number of subscribers instead of a user's data rate. Using this approach, it was found that especially the power consumption of high data rate applications (file download and video streaming) is much lower than reported before \cite{Malmodin20}. 
 
As a consequence, in Section~\ref{sec:local}, we construct our power model based on a separation of a bitrate-dependent variable power and the constant offset power.

\subsection{Content Delivery Networks}
\label{secsec:globCDN}
Certain online video services are operated using a so-called content delivery network (CDN) \cite{Bianco16}. 
The concept of a CDN considers a network of spatially distributed DCs, which communicate over the Internet. A CDN is especially useful if content shall be stored close to the end users, for example, in on-demand video streaming \cite{Bianco16}. A general assumption for CDNs is that there is a main server storing all videos and that there are additional surrogate servers hierarchically structured, which store copies of the most requested videos \cite{Bianco16}. Due to additional management and regular updates of stored videos, such CDNs consume additional energy, which is discussed in detail in \cite{Bianco16,Boscovic11,Goudarzi20}.

\subsection{End Users}
\label{secsec:globUT}
The energy consumed by end users must be handled differently than the energy consumed by DCs and transmission networks. The reason is that DCs and transmission networks usually serve a very large number of end users (several thousands of end users for mobile networks in cities or up to millions of end users for Internet exchange nodes), where in contrast, end-user devices usually only serve a small number of users (one to a couple of end users), but several millions of end-user devices must be considered.
 
Some overall estimates for the worldwide energy consumption of end-user devices are given in \cite{ShiftFull19, Malmodin20}. In these studies, the main devices are displaying devices such as smartphones, tablet PCs, and television (TV) sets including the display. Utilities such as routers and local network switches, which were already discussed in Subsection~\ref{secsec:globNW}, are also considered. Typically, it is taken into account that routers are running continuously with approximately constant power consumption, whereas displaying devices only consume power during video streaming \cite{Malmodin20}. Typical energy and power consumption values for these devices can be found in device manuals and were analyzed in detail in different studies \cite{Malmodin20, Herglotz20, Sadasivan16, Belkhir18,Herglotz22a}. 

Earlier studies found that the total energy consumption of all end-user devices performing online video tasks is approximately the same as that of all DCs and all transmission networks individually \cite{ShiftFull19, Hintemann16}. However, recent studies suggest that in fact, end-user devices contribute most to the overall power consumption during streaming \cite{Malmodin20}, where it is taken into account that cloud services support many more applications than just online video applications. Hence, the use of power efficient end-user devices, e.g., the use of smartphones and tablet PCs instead of large TV screens or desktop PCs is highly beneficial for saving power \cite{Malmodin20}.

\section{Energy Consumption Modeling}
\label{sec:local}

In this section, we construct an energy model by beginning with the worldwide energy consumption $\boldsymbol{E}$ caused by online video applications per year. 
The model covers the major online video services including applications such as 
\begin{itemize}
\item on-demand streaming,
\item teleconferencing,
\item remote Desktop,
\item peer-to-peer (P2P),
\item content delivery networks (CDN),
\item surveillance,
\item Internet protocol television (IPTV),
\item social media, 
\item messaging,
\item cloud backup.
\end{itemize}	
Note that we explicitly only consider the energy consumption of services and devices related to direct and live online streaming. That means that regarding the recording of videos, we consider the case of a single capturing device (i.e., end-user recording for teleconferencing or social networks). In contrast, we do not consider complex systems for live TV broadcast or movie production. Furthermore, we do not take into account the energy consumed for the  production and disposal of devices or for further related activities such as the standardization efforts needed for the establishment of transmission networks and video codecs.  

In Subsection~\ref{secsec:globalEnergy}, we define the model at the global level and assign all devices needed during the use of online video services to the global online video system, which consumes the energy $\boldsymbol{E}$ every year. The energy on this global level is then split up into the energy consumption caused by the online video services mentioned above, which corresponds to the viewpoint of a system engineer of a single service. Afterwards, in Subsections~\ref{secsec:EUD} to \ref{secsec:NW}, we develop more detailed energy models by considering the end-user terminals, the networks, and the provider's data centers separately. {\color[rgb]{0,0,0} Finally, we discuss how the model can be validated in Subsection~\ref{secsec:valid}.}

\subsection{Model Definition}
\label{secsec:globalEnergy}

As practically, it is difficult to measure and describe the global energy consumption $\boldsymbol{E}$ caused by all online video services, we break this energy up into parts that we can estimate using values reported in the literature. To this end, we refer to the reports released by the video service providers, which often provide statistical data on the use of their services as, e.g., listed in \cite{Malmodin20}. The system engineer can then consider one service at a time. Hence, we split up the worldwide energy consumption as follows
\begin{equation}
\boldsymbol{E} = \sum_{s\in S}{E_s}, 
\label{eq:glonEnergy}
\end{equation}
where $S$ is the complete set of online video services and $s$ the index of a single service. The energy $E_s$ is the energy consumption caused by the $s$-th service per year. For example, this energy represents the total energy consumption caused by an on-demand streaming service, a teleconferencing service, or a social media platform. 

Then, the system engineer can further decompose the service-level energy as follows
\begin{equation}
E_s 
= E_{s,\mathrm{UT}} + E_{s,\mathrm{VP}} + E_{s,\mathrm{NW}},
\label{eq:overallPowerModel}
\end{equation}
where $E_{s,\mathrm{UT}}$ is the energy demand of all end-user terminals using the $s$-th video service, $E_{s,\mathrm{VP}}$ is the yearly energy consumed by the  service's DCs and servers, and $E_{s,\mathrm{NW}}$ the energy required for the  transmission of the videos via networks.

\subsection{End-user Devices}
\label{secsec:EUD}
In this subsection, we consider the total yearly energy consumption of all end-user devices such as tablet PCs, TVs, smartphones, and desktop PCs including their {\color[rgb]{0,0,0}displays}. 
We sum up over all $D_{s}$ end-user devices, which make use of service $s$, as follows
\begin{equation}
E_{s,\mathrm{UT}} = \sum_{d\in D_{s}}   E_{s,\mathrm{UT},d},  
\label{eq:serviceWiseUTEnergy}
\end{equation}
where $E_{s,\mathrm{UT},d}$ denotes the energy consumption of the $d$-th device for the $s$-th service. Next, note that each device can request multiple streaming events, which are summed up in the set of requests $R_{s,\mathrm{UT},d}$. The total energy consumption of device $d$ can then be calculated by summing up the energy consumption of all requests $r$ as follows
\begin{equation}
E_{s,\mathrm{UT},d} = \sum_{r\in R_{s,\mathrm{UT},d}}   E_{s,\mathrm{UT},d,r}. 
\label{eq:deviceWiseUTEnergy}
\end{equation}
Finally, each request $r$ is characterized by the direction of streaming. This means that video streaming can be performed either by download and playback (Rx), by capture and upload (Tx), or both can be done at the same time. To cover these possibilities, we propose to compute the request-wise energy consumption based on \cite{Malmodin20,Herglotz20,Carroll13} as
\begin{equation}
\label{eq:eventWiseUTEnergy}
E_{s,\mathrm{UT},d,r} = \left( p_{s,\mathrm{UT},d,r,0}+ p_{s,\mathrm{UT},d,r,\mathrm{Rx}}+p_{s,\mathrm{UT},d,r,\mathrm{Tx}}\right) \cdot t_r,
\end{equation}
where we multiply the total power consumption of the device with the duration of the requested streaming event $t_r$. The total power consumption is the sum of the constant offset power consumption $p_{s,\mathrm{UT},d,r,0}$, the receiving and playback power $p_{s,\mathrm{UT},d,r,\mathrm{Rx}}$, and the capturing and upload power consumption $p_{s,\mathrm{UT},d,r,\mathrm{Tx}}$. 
If only one direction (Rx or Tx) is active, the power for transmission in the opposite direction is set to zero. 

First, we discuss the constant offset power $p_{s,\mathrm{UT},d,r,0}$ for a single streaming request. This power mainly depends on the type of device and its components. For example, a smartphone usually demands only a few Watts \cite{Herglotz20,Carroll13} due to its energy efficient architecture. Modern tablet PCs require about $10\,$W to $30$\,W \cite{Malmodin20,Herglotz22a}, the power consumption of TV sets mainly depends on the display technology (liquid-crystal display (LCD) or organic light-emitting diode (OLED) display) and their screen size ($50\,$W to $200$\,W) \cite{Sadasivan16}, and the power consumption of desktop PCs depends on their hardware configuration and the size and type of the monitor (power consumption values similar to TVs) \cite{Ong14}.   

For the receiving and playback power $p_{s,\mathrm{UT},d,r,\mathrm{Rx}}$, the effect of certain video parameters on the power was studied in various papers. For example, a very detailed study was performed on the influence of video parameters such as bitrate, resolution, frame rate, {\color[rgb]{0,0,0}and video codec} for smartphones \cite{Herglotz20}. 
First, it was found that the power consumption depends on the smartphone model and the chosen application, which indicates that also the software implementation (in terms of the written code) affects the power consumption. Moreover, the choice of the network connection (wireless, wired) was shown to influence the power consumption of the device. In general, it was found that during online video streaming, up to $50\%$ of the power consumption (including the constant offset power) can be controlled by manipulating the aforementioned parameters \cite{Herglotz20}. Similar results could be observed for laptops and desktop PCs \cite{Herglotz22a}, however, the constant offset power had a higher influence in general. Unfortunately, to the best of the authors' knowledge, similar studies on power consumption are not available for TVs. However, due to the similarity of the technologies used, we expect that the same kind of dependencies can be observed. A detailed analysis of this conjecture is an interesting topic for future research. 

Other studies focused on specific subprocesses in end-user devices such as video decoding in hardware and software \cite{Khernache21,Mallikarachchi16b,Herglotz18a,Correa18,Herglotz18,Kraenzler19}, where also devices such as desktop PCs were considered. The influence of streaming protocols and different kinds of networks (wired/wireless) was studied in \cite{Zhang18,Zhang16}. Furthermore, it was shown that for LCDs, the backlight brightness can be adjusted to the brightness of the video content without sacrificing image quality, which can save power, especially for large screens \cite{Liu16}.

The capturing and uploading power consumption $p_{s,\mathrm{UT},d,r,\mathrm{Tx}}$ was discussed in detail in \cite{Feng05, Kleihorst01, Saffari19,Hao11,Lee12}. However, to the best of the authors' knowledge, no dedicated power or energy models were developed yet, which is another interesting topic for future research. 

Table~\ref{tab:smart} shows specific power values for different smartphones from three different studies \cite{Carroll13,Herglotz20,Hao11}. Furthermore, Table~\ref{tab:otherUTs} lists some rough power estimates for other devices (tablet PCs, laptops, TVs, and desktop PCs) that were reported in \cite{Malmodin20}, where we could only find data for the constant offset power $p_{s,\mathrm{UT},d,r,0}$. 
\begin{table}[t]
\caption{Exemplary modeling parameters for various smartphones. 
The first device was measured during local playback and capture, the second device and the third device were connected using a Wi-Fi connection. If parameter ranges are displayed, the true value depends on video parameters such as the bitrate. The asterisk is a replacement for the index $*\,\widehat{=}\, ``s,\mathrm{UT},d,r\text{''}$.} 
\label{tab:smart}
\vspace{-.4cm}
\begin{center}
\def\arraystretch{1.5}
\footnotesize
{\begin{tabular}{r|r|r|r}
\hline
 & Samsung Galaxy S III 
  & Fairphone 2 & HTC G1\\ 
 & \cite{Carroll13}, 2013 & \cite{Herglotz20}, 2020 & \cite{Hao11}, 2011\\
\hline
$p_{*,0}$ & $805\,$mW & $900\,$mW  & $319\,$mW \\
$p_{*,\mathrm{Rx}}$&  $279\,$ - $1{,}524\,$mW & up to $2.21\,$W  & -\\
$p_{*,\mathrm{Tx}}$ & min. $1.809\,$W  & - & min. $1.05\,$W\\
 \hline
 \end{tabular}}
\end{center}
\end{table}
\begin{table}[t]
\caption{Rough modeling parameters for further end-user devices as reported in \cite{Malmodin20}. The asterisk is a replacement for the index $*\,\widehat{=}\, ``s,\mathrm{UT},d,r\text{''}$.} 
\label{tab:otherUTs}
\vspace{-.4cm}
\begin{center}
\def\arraystretch{1.5}
\footnotesize
{\begin{tabular}{r|c|c|c|c}
\hline
 & Tablet & Laptop & TV & PC\\ 
\hline
$p_{*,0}$ & $10\,$W & $20\,$W & $100\,$W & $100\,$W  \\
 \hline
 \end{tabular}}
\end{center}
\end{table}
Values for the other parameters are not given because they were not provided in the respective literature. It is further important to note that some of the values provided in Tables~\ref{tab:smart} and \ref{tab:otherUTs} are derived from values reported in the literature. For example, the constant offset power  $p_{s,\mathrm{UT},d,r,0}=805\,$mW in the left column of Table~\ref{tab:smart} corresponds to the idle power in airplane mode reported in \cite{Carroll13}. To obtain the receiving power, we then subtracted this idle power from the reported overall video playback power as $p_{s,\mathrm{UT},d,r,\mathrm{Rx}} = 1084\,\mathrm{mW}-805\,\mathrm{mW} = 279\,\mathrm{mW}$ (for the low-end video power). The other values were derived analogously. 

The values in Table~\ref{tab:smart} show that the power consumption depends significantly on the device. Furthermore, various parameters are shown in a certain range, as reported in the referenced studies \cite{Carroll13,Herglotz20,Hao11}. The variations can be attributed to the video parameter settings (bitrate, frame rate, etc.) and other configuration options. 

%
%

%
%
%

%
%

{\color[rgb]{0,0,0} 
For more accurate modeling of the overall end-user energy consumption, one would have to consider the dependency of the powers, i.e., $p_{s,\mathrm{UT},d,r,0}$, $p_{s,\mathrm{UT},d,r,\mathrm{Rx}}$, and $p_{s,\mathrm{UT},d,r,\mathrm{Tx}}$, on additional parameters such as video properties and screen brightness settings. This relation was studied extensively in the literature \cite{Herglotz20,Mallikarachchi16b,Herglotz18a,Correa18,Herglotz18,Kraenzler19,Zhang18,Zhang16,Liu16,Khernache21,Herglotz22a} and could be included in the model proposed here. However, in this paper, we rather focus on a realistic range of power consumption values, which is sufficient for modeling the impact of high-level design choices of the system engineer in Section~\ref{sec:energyEfficiency}. A more detailed modeling approach could be developed in future research. 
 }

\subsection{Video Providers}
\label{secsec:Providers}

To model the yearly provider-side energy consumption of a single service, 
{\color[rgb]{0,0,0}
we can choose either a bottom-up or a top-down approach. For the top-down approach, one could use statistical information of the yearly energy consumption of all worldwide data centers (DCs) and their corresponding share of workload for online video systems \cite{Hintemann16,statistaDataCenters}. For general-purpose DCs in the cloud, a large amount of research has been performed, which analyzes the overall power consumption in detail  \cite{Zhou18,Heller10,Tang18}. As a drawback, the energy values obtained in this manner provide no information on the power consumption related to the specific devices and the exact algorithms that were deployed in the DC. 

Hence, in this paper, we follow the bottom-up approach and}
consider single devices in DCs as well as CDNs{\color[rgb]{0,0,0}, which are servers. Hence, we split up the energy consumption of DCs into the energy consumption of single servers}. For DCs explicitly targeting online video applications, the literature on energy modeling focuses on CDNs \cite{Bianco16,Boscovic11,Chan11,Goudarzi20}, which we take as the basis for our considerations. As a consequence, we model the yearly energy consumption of a CDN as the sum of the energy consumption of all related servers as
\begin{equation}
E_{s,\mathrm{VP}} = \sum_{\sigma\in \mathit{\Sigma_s}} \eta_\sigma E_{s,\mathrm{VP},\sigma}, 
\label{eq:VP_serverSum}
\end{equation}
where $\sigma$ is the index for a single server, $\mathit{\Sigma}_s$ is the set of servers used by the $s$-th service, $\eta_\sigma$ is the PUE of the $\sigma$-th server, and $E_{s,\mathrm{VP},\sigma}$ denotes the yearly energy consumption of the $\sigma$-th server. 

\begin{figure}[t]
\centering
\psfrag{I}[cl][l]{\hspace{-0.7cm}Input videos}
\psfrag{V}[c][c]{Receive videos }
\psfrag{R}[c][c]{}
\psfrag{T}[l][l]{Transcode}
\psfrag{E}[l][l]{videos}
\psfrag{W}[l][l]{}
\psfrag{S}[r][r]{Copy to surrogate servers}
\psfrag{U}[c][c]{Send videos to }
\psfrag{X}[c][c]{end users}
\psfrag{C}[c][c]{CDN with $\mathit \Sigma_s$ servers}
\includegraphics[width=.49\textwidth]{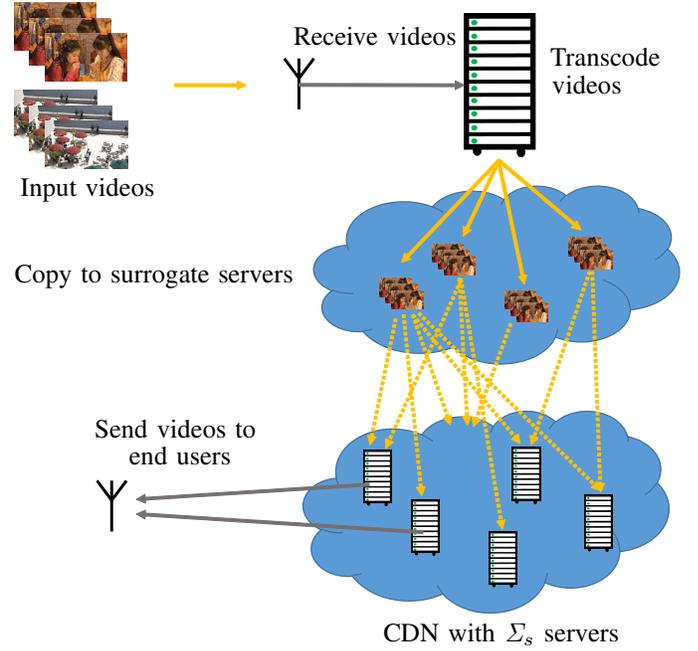} 
\caption{Tasks of a provider to operate an online video service.  }
\label{fig:provider}
\end{figure}

Based on the modeling of a CDN presented in \cite{Bianco16}, the tasks of the servers are defined as follows (see Fig.~\ref{fig:provider}): First, the server receives input videos (a movie, a live stream from, e.g., a football match, a video from a user to share via social networks, or a live teleconferencing stream). Depending on the type of service, it is then transcoded to different codecs or bitrates to generate additional videos. Afterwards, in the case of on-demand video, the videos are stored on one server (e.g., in a DC) or on multiple servers in a CDN. In the case of a CDN, the video is transmitted to a set of surrogate servers, which store copies of the video to reduce the spatial distance to the end user \cite{Bianco16}. Finally, the videos are transmitted to the end user in a single-cast or multi-cast fashion triggered by user requests. Note that depending on the video streaming application, some of these tasks may not be relevant. For example, the provider of a peer-to-peer (P2P) service (e.g., video conferencing) does not have to receive, transcode, and send the video stream itself. Hence, it is sufficient that the servers manage the user accounts and the connections between the peers.  

Consequently, we model the overall yearly energy consumption of a server based on \cite{Bianco16} as follows
\begin{align}
E_{s,\mathrm{VP},\sigma} = & E_{s,\mathrm{VP},\sigma,0}+  E_{s,\mathrm{VP},\sigma,\mathrm{Rx}} + E_{s,\mathrm{VP},\sigma,\mathrm{trans}} \label{eq:VP_serverParts} \\ & + E_{s,\mathrm{VP},\sigma,\mathrm{copy}} + E_{s,\mathrm{VP},\sigma,\mathrm{store}} + E_{s,\mathrm{VP},\sigma,\mathrm{Tx}},
\notag
\end{align}
where $E_{s,\mathrm{VP},\sigma,0}$ refers to a constant offset energy consumption in idle mode, $E_{s,\mathrm{VP},\sigma,\mathrm{Rx}}$ denotes the energy consumption for receiving videos, $E_{s,\mathrm{VP},\sigma,\mathrm{trans}}$ is the energy consumption for transcoding videos, $E_{s,\mathrm{VP},\sigma,\mathrm{copy}}$ represents the energy consumed for sending copies to surrogate servers, $E_{s,\mathrm{VP},\sigma,\mathrm{store}}$ is the energy consumed for storing videos, and $E_{s,\mathrm{VP},\sigma,\mathrm{Tx}}$ denotes the energy consumption for sending (transmitting) the videos to the end users.

First, we consider the server-side energy needed to send videos to the end users. We adopt the model proposed in \cite{Bianco16} that is given by 
\begin{equation}
E_{s,\mathrm{VP},\sigma,\mathrm{Tx}} = \sum_{r\in R_{s,\mathrm{VP},\sigma,\mathrm{Tx}}}  B_r \cdot e_{s,\mathrm{VP},\sigma,\mathrm{send}}, 
\label{eq:VP_Tx}
\end{equation}
where $r$ indicates a requested video from the set of video requests $R_{s,\mathrm{VP},\sigma,\mathrm{Tx}}$, which includes all requests directed to the $\sigma$-th server.  $B_r$ is the size of the requested video in bits and $e_{s,\mathrm{VP},\sigma,\mathrm{send}}$ in $\frac{\mathrm{J}}{\mathrm{bit}}$ is the energy consumption needed to send a single bit to an end user. Similarly, the energy needed to send copies of a video to surrogate servers is given by 
\begin{equation}
E_{s,\mathrm{VP},\sigma,\mathrm{copy}} = \sum_{r\in R_{s,\mathrm{VP},\sigma,\mathrm{copy}}}  B_r \cdot e_{s,\mathrm{VP},\sigma,\mathrm{send}}, 
\label{eq:VP_copy}
\end{equation}
where $r$ indicates a requested copying event of a video from the set of copying events $R_{s,\mathrm{VP},\sigma,\mathrm{copy}}$ on the $\sigma$-th server. 
For simplicity, we assume that the energy needed to send data to an end user $e_{s,\mathrm{VP},\sigma,\mathrm{send}}$ from Eq.~\eqref{eq:VP_copy} is the same as the energy needed to send data to a surrogate server. 

For receiving videos, we model the energy in a similar manner \cite{Bianco16,Chan11} such that  
\begin{equation}
E_{s,\mathrm{VP},\sigma,\mathrm{Rx}} = \sum_{r\in R_{s,\mathrm{VP},\sigma,\mathrm{Rx}}}  B_r \cdot e_{s,\mathrm{VP},\sigma,\mathrm{Rx}}, 
\label{eq:VP_Rx}
\end{equation}
where $r$ indicates a request to receive a video from the set of requests $R_{s,\mathrm{VP},\sigma,\mathrm{Rx}}$ on the $\sigma$-th server and 
$e_{s,\mathrm{VP},\sigma,\mathrm{Rx}}$ is the energy consumption per received bit. 

For the transcoding energy consumption, we take into account that in many applications, a single input video is transcoded to multiple videos with multiple codecs, multiple bitrates, and with varying spatial as well as temporal resolutions to account for differences in user requirements and bandwidth restrictions \cite{Wu20}. The corresponding energy consumption is thus split into decoding energy consumption and encoding energy consumption as follows
\begin{align}
E_{s,\mathrm{VP},\sigma,\mathrm{trans}} = & \sum_{r\in R_{s,\mathrm{VP},\sigma,\mathrm{trans}}}  \Big(  e_{s,\mathrm{VP},\sigma,\mathrm{trans},r,{\mathrm{dec}}}   \\ &  +\sum_{v\in V_{s,\mathrm{VP},\sigma,\mathrm{trans},r}} e_{s,\mathrm{VP},\sigma,\mathrm{trans,}r,{\mathrm{enc},v}} \Big)\notag, 
\end{align}
where decoding of the video based on the $r$-th transcoding request from the set of transcoding requests $R_{s,\mathrm{VP},\sigma,\mathrm{trans}}$ is performed once. For each transcoding request, encoding can be performed multiple times, which is modeled by the sum over the set of output videos $V_{s,\mathrm{VP},\sigma,\mathrm{trans},r}$  indexed by $v$. $e_{s,\mathrm{VP},\sigma,\mathrm{trans},r,{\mathrm{dec}}}$ is the decoding energy consumption of the $r$-th video and $e_{s,\mathrm{VP},\sigma,\mathrm{trans,}r,{\mathrm{enc},v}}$ is the encoding energy consumption of the $v$-th video for the $r$-th transcoding request. 

The modeling of the decoding energy consumption was discussed in detail in  \cite{Mallikarachchi16b,Herglotz18a,Correa18,Herglotz18,Kraenzler19,Herglotz15c}. It was shown that it depends on a multitude of video parameters such as the bitrate, the frame rate, the resolution, the type of video codec, and the software or hardware implementation, whichever is used, respectively. Explicit energy consumption values depending on these parameters can be derived from, e.g.,  \cite{Herglotz18a,Herglotz15c}. {\color[rgb]{0,0,0} Still, it was also shown that an energy model solely based on the decoding time and the mean decoding power can be sufficient for accurate modeling \cite{Herglotz15a}. We adopt this modeling approach here to be consistent with the definition of the energy consumption of end-user devices (see Section~\ref{secsec:EUD}). Hence, }the decoding energy consumption of a single video is given by
\begin{equation}
e_{s,\mathrm{VP},\sigma,\mathrm{trans},r,{\mathrm{dec}}} = p_{s,\mathrm{VP},\sigma,\mathrm{trans},r,{\mathrm{dec}}}\cdot t_r,
\end{equation}
where $t_r$ is the duration of the video corresponding to the $r$-th transcoding request. Here, $p_{s,\mathrm{VP},\sigma,\mathrm{trans},r,{\mathrm{dec}}}$ in $\frac{\mathrm{J}}{\mathrm{s_\mathrm{video}}}$ refers to the energy consumption needed to decode one second of video content, which is given in the unit $\mathrm{s}_\mathrm{video}$ for the $r$-th request on the $\sigma$-th server. Note that in the case of real-time decoding, which is not required in DCs, this metric is the same as the decoding power consumption in Watts.  

Considering the encoding energy consumption, different models are presented in \cite{Liang04,Kim11,Lu04}, which show that it depends on the same parameters (e.g., resolution, frame rate) as the decoding energy consumption. Further works target the optimization of the encoder energy consumption \cite{He05, Ahmad09, Mercat17, Saab14, Penny16}, where additional parameters influencing the energy consumption have been derived. Next to the parameters mentioned for decoding, these are mainly the encoder configurations that have a significant influence on the encoding complexity. 
Examples are the bit stream configuration such as all-intra coding or random-access coding \cite{Bossen13} and the choice of the target bitrate or the target visual quality using a quantization parameter (QP) \cite{Bossen12}. Additionally, many encoders (e.g., x264 \cite{x264} or x265 \cite{x265}) can be tuned, where so-called presets and other encoding parameters can be chosen to control the encoding speed \cite{Wu20} or the target compression performance. Note that these parameters can also affect the power consumption of recording videos on end-user devices \eqref{eq:eventWiseUTEnergy}, as the encoding of videos is usually done simultaneously with capturing. {\color[rgb]{0,0,0}To be consistent with the model for the decoding energy}, we model the encoding energy consumption for a single video as
\begin{equation}
e_{s,\mathrm{VP},\sigma,\mathrm{trans},r,{\mathrm{enc}},v} = p_{s,\mathrm{VP},\sigma,\mathrm{trans},r,{\mathrm{enc}},v}\cdot t_r,
\end{equation}
where $p_{s,\mathrm{VP},\sigma,\mathrm{trans},r,{\mathrm{enc}},v}$ denotes the encoding energy per video second of the $v$-th video to be encoded. In \cite{Ramasubbu22a}, it is shown that this kind of energy model is sufficiently accurate (estimation error below $8\%$ on average).

After transcoding, the videos need to be stored on the servers for permanent availability. The corresponding energy consumption was discussed and modeled in \cite{Bianco16} as follows
\begin{equation}
E_{s,\mathrm{VP},\sigma,\mathrm{store}} = e_{s,\mathrm{VP},\sigma,\mathrm{store}}\cdot \sum_{v\in V_{s,\mathrm{VP},\sigma,\mathrm{store}}} B_v,
\end{equation} 
where $e_{s,\mathrm{VP},\sigma,\mathrm{store}}$ in $\frac{\mathrm{J}}{\mathrm{bit}\cdot\mathrm{year}}$ is the energy consumed for storing a single bit on the $\sigma$-th server for one year. $v$ indicates one video of the set of videos $V_{s,\mathrm{VP},\sigma,\mathrm{store}}$ stored on the $\sigma$-th server. 

Table~\ref{tab:energyParamsVP} lists energy consumption values taken from the literature. 
\begin{table}[t]
\caption{Energy parameters for video providers. The encoding power consumption and the decoding power consumption are listed for an HD video at $30$\,fps, and a bitrate of $2$\,$\frac{\mathrm{Mbit}}{\mathrm{s}}$. }
\label{tab:energyParamsVP}
\vspace{-.4cm}
\def\arraystretch{1.5}
\begin{center}
\footnotesize
{\begin{tabular}{r|r|r}
\hline
 Parameter & Value  & Source \\
\hline
$\eta$ & $1.08$ & \cite{Uzaman19}\\
$E_{s,\mathrm{VP},\sigma,0}$ & $127\,\mathrm{kWh}-5.57\,$MWh & \cite{Dayarathna16}\\
$p_{s,\mathrm{VP},\sigma,\mathrm{trans,}r,{\mathrm{enc}}}$ & $200\,\frac{\mathrm{mJ}}{\mathrm{s}_\mathrm{video}}-90\,\frac{\mathrm{kJ}}{\mathrm{s}_\mathrm{video}}$    & \cite{Zhang18HWEnc,Penny16}\\
$p_{s,\mathrm{VP},\sigma,\mathrm{trans,}v,{\mathrm{dec}}}$  & $719\,\frac{\mathrm{mJ}}{\mathrm{s}_\mathrm{video}}-24.45\,\frac{\mathrm{J}}{\mathrm{s}_\mathrm{video}}$   & \cite{Herglotz18a,Herglotz15c}\\
$e_{s,\mathrm{VP},\sigma,\mathrm{store}}$ & $0.59\,\frac{\mathrm{Wh}}{\mathrm{MByte}}$  per year  & \cite{Bianco16}\\
 $ e_{s,\mathrm{VP},\sigma,\mathrm{send}} = e_{s,\mathrm{VP},\sigma,\mathrm{Rx}}$ & $0.624\,\frac{\mathrm{mWh}}{\mathrm{MByte}}$    & \cite{Bianco16}\\
 \hline
 \end{tabular}}
\end{center}
\end{table}
For the PUE $\eta$, we choose a constant value for all servers which was reported for a social network in \cite{Uzaman19}. 
For the encoding energy consumption, we provide a range of potential values because it highly depends on the used technology and the implementation of the encoder. It is worth mentioning that the low value for encoding one hour of video ($200\,$mWh) is more than five orders of magnitude smaller than the high value ($90\,$kWh). The reason for this extreme difference is that for the low end, a highly optimized hardware encoder chip customized for low power applications was used, whereas for the high end, an extremely complex software implementation targeting a maximum compression performance was employed. A similar range is given for the decoding power consumption, where the lower value is reported for hardware decoding and the higher value for HEVC software decoding, both on an evaluation platform.


{\color[rgb]{0,0,0}
Finally, we note that similar to end-user devices, also the energy consumption values in DCs  differ significantly depending on the deployed hardware components (CPUs, GPUs, FPGAs) and the video properties. For more accurate models, these factors need to be taken into account, especially as energy values can differ by several orders of magnitude (e.g., those for video encoding as mentioned above). However, in this paper, we focus on the reported range of values instead of a more accurate modeling approach to obtain upper and lower bounds on the total energy consumption. A more accurate modeling approach and its implications can be considered in future work. 
}

\subsection{Transmission Networks}
\label{secsec:NW}

For the yearly energy consumption related to transmission, one approach is to separately consider all network nodes that a certain video streaming service exploits, as was proposed in \cite{Coroama13}. Network nodes include the main network components such as Internet exchange points (IXPs), local area network (LAN) routers, access networks, switches, repeaters, home routers, mobile base stations (4G/5G), etc. A thorough analysis of the types of network nodes and their power characteristics is beyond the scope of this work but can be found in the literature \cite{Coroama13,Malmodin14,Malmodin20}.  

Similar to end-user devices, the node-wise energy consumption again depends on the hardware of the node, the configuration, the software, the software configuration, and the type of network (wired, wireless). Estimations and analyses on the energy consumption of such nodes can be found in the literature \cite{Malmodin20, Coroama13, Zhang19, Malmodin14}, but are not further discussed here. 

For energy consumption modeling, we adopt an approach from a recent study \cite{Malmodin20}, which proposed a simplified view based on transmission requests. In this approach, the network nodes are not considered separately, but the overall energy consumption of all nodes concerned by a single streaming request is used as a basis.  As this approach can elegantly be mapped to the modeling of end-user devices and provider-side servers discussed above, we adopt it for modeling the energy consumption of the network as 
\begin{equation}
E_{s,\mathrm{NW}} = E_{s,\mathrm{NW,UT}} + E_{s,\mathrm{NW,CDN}},
\end{equation}
where $E_{s,\mathrm{NW,UT}}$ is the energy consumption of all network nodes caused by end-user requests and $E_{s,\mathrm{NW,CDN}}$ is the energy consumption of all network nodes caused by copying of videos to surrogate servers in a CDN. 

The energy consumption caused by end-user requests is then modeled based on \cite{Malmodin20} as follows
\begin{align}
&E_{s,\mathrm{NW,UT}} =\label{eq:NWUT} \\
& \sum_{d\in D_{s}}\left[ \sum_{r\in R_{s,\mathrm{UT,}d}} \hspace{-.4cm}\left(p_{s,\mathrm{NW,UT},d,r,0} + p_{s,\mathrm{NW,UT},d,r,\mathrm{rate}}\cdot b_r \right) \cdot t_r \right],\notag
\end{align}
where we sum over all streaming requests $r$ in the set $R_{s,\mathrm{UT,}d}$ from all devices $d$ in the set $D_{s}$, where the same sets as already defined in Eqs.  \eqref{eq:serviceWiseUTEnergy} and \eqref{eq:deviceWiseUTEnergy} are reused, respectively. The energy of a single request is then given as the product of the duration of the request $t_r$ and the transmission power, which is the sum of a constant offset power $p_{s,\mathrm{NW},d,r,0}$ and a variable power $p_{s,\mathrm{NW},d,r,\mathrm{rate}}$ in $\left[\frac{\mathrm{W}}{\mathrm{bps}}\right]$. The constant offset power corresponds to the idle power of network devices 
and the variable power depends on the bitrate $b_r$ of the requested video \cite{Malmodin20}. 

Likewise, the energy consumption of data transmission caused by copying of videos in CDNs can be modeled as \cite{Malmodin20,Bianco16}
\begin{align}
&E_{s,\mathrm{NW,CDN}} =\label{eq:NWVP} \\
& \sum_{\sigma\in \mathit{\Sigma}_{s}}\left[ \sum_{r\in R_{s,\mathrm{VP,}\sigma,\mathrm{copy}}} \hspace{-.7cm}\left(p_{s,\mathrm{NW,VP},\sigma,r,0} + 
p_{s,\mathrm{NW,VP},\sigma,r,\mathrm{rate}}\cdot b_r \right)
 \cdot t_r \right],\notag
\end{align}
where we sum over all copy requests $r$ in the set $R_{s,\mathrm{VP},\sigma}$ and all servers $\sigma$ in the set $\mathit{\Sigma}_{s}$, which are the same sets as defined in Eqs.  \eqref{eq:VP_serverSum} and \eqref{eq:VP_copy}, respectively.

Power values for modeling the network power consumption are listed in Table~\ref{tab:energyParamsNW}. 
\begin{table}[t]
\caption{Exemplary power parameters for data transmission. All values are taken from \cite{Malmodin20}.  }
\label{tab:energyParamsNW}
\vspace{-.4cm}
\begin{center}
\def\arraystretch{1.5}
\footnotesize
{\begin{tabular}{r|r|r}
\hline
 & \textbf{Fixed BB access } &\textbf{Mobile BB Access (4G)} \\
\hline
$p_{s,\mathrm{NW,UT},d,r,0}$ &  $1.5\,\mathrm{W}$ & $0.2\,\mathrm{W}$   \\
$p_{s,\mathrm{NW,UT},d,r,\mathrm{rate}}$ & $0.03\,\frac{\mathrm{W}}{\mathrm{Mbps}}$    & $0.03\,\frac{\mathrm{W}}{\mathrm{Mbps}}$\\
 \hline
 \hline
  & \textbf{Optical Link} & \textbf{Atlantic cable }\\
  \hline
  $p_{s,\mathrm{NW,VP},\sigma,r,\mathrm{rate}}$ & $0.1\,\frac{\mathrm{W}}{\mathrm{Mbps}}$ & $0.05\,\frac{\mathrm{W}}{\mathrm{Mbps}}$\\
  \hline
 \end{tabular}}
\end{center}
\end{table}
The top two rows report values for two types of network connections for end-user devices, namely a fixed broadband (BB) access network and mobile BB access network. The bottom row reports power values for high capacity optical links and an Atlantic optical submarine cable for long-distance data transmission. All values are taken from \cite{Malmodin20}, where it is reported that the distance of data transmission can be neglected.

{\color[rgb]{0,0,0}
\subsection{Model Validation}
\label{secsec:valid}
The model developed in this paper is difficult to validate in general. To still obtain acceptable estimates, we constructed the proposed energy consumption model based on existing models for the subsystems, which were reported in the literature and that have already been validated. Nevertheless, considering the rapid technological progress, the reported model parameters could already be outdated. Still, the estimated energy values for the overall energy consumption of video services, which are derived in the next section, provide first insights into the importance of this high-level modeling approach. 

As another validation step, in future work, one could set up a complex network of measurement equipment to measure the power consumption of the various end-user devices, network nodes, and the DCs performing online video tasks. Resulting model parameters can then be used to update the overall energy consumption estimates reported in this paper.  
The same holds for the energy consumption of transmission via the global Internet, where other kinds of data are transmitted at the same time. 

}

\section{Model Implications}
\label{sec:energyEfficiency}
In this section, we will discuss the implications of the model presented in the previous section and some related lessons that we can learn. To this end, we take the point of view of a system engineer whose task it is to set up an online video platform in the most sustainable, i.e., energy-efficient manner. At first, in Subsection~\ref{secsec:serviceEnergy}, we show how the proposed model can help the system engineer to determine the most important energy consumers, which strongly depends on the type of online video service. Subsequently, in Subsection~\ref{secsec:singleVideo}, we discuss a more detailed example for a single video and show that considering energy consumption when designing an online video service can lead to significant energy savings. Finally, in Subsection~\ref{secsec:approaches}, we unveil future research directions for improving the energy efficiency of online video. 

{\color[rgb]{0,0,0}
Note that the considerations in the next subsections rely on parameter values that may not be fully representative of the state-of-the-art hardware used in practice. As a consequence, the resulting modeling values may not represent the true energy consumption of the services. However, from the system engineer's point of view, the results show clearly which hardware components and parameters are most important to address to increase the energy efficiency of an online video system, which underlines the usefulness of our proposed model. 
}

\subsection{The Energy Consumption of Online Video Services}
\label{secsec:serviceEnergy}
In this section, using the proposed model, we perform a thought experiment by estimating and comparing the overall energy consumption of different online video services. The parameters, i.e., the number of requests and the corresponding set of devices, are selected based on the values of an actual on-demand platform \cite{Malmodin20}. To allow comparisons with the energy consumption of other kinds of services, we use similar values for other services. Hence, the adopted values do not represent actual services, but help to understand the energy consumption of the most important components of online video systems.

Assume that you are a system engineer who is asked to construct an online video system serving $100$ million end users that employ $\left|D_s\right|=100\cdot 10^6$  distinct devices. 
Each end user makes use of the service one hour per day on average. Each video shall be coded with high visual quality such that the bitrate is set to a value between $b_r=2\,\frac{\mathrm{Mbit}}{\mathrm{s}}$ and $b_r=10\,\frac{\mathrm{Mbit}}{\mathrm{s}}$ \cite{Herglotz20}, where for simplicity, we adopt a mean value of $b_r=5\,\frac{\mathrm{Mbit}}{\mathrm{s}}$ in the following. Furthermore, the system engineer can make use of $\left|\mathit{\Sigma}_s\right|=1000$ servers worldwide to construct and operate the service using, e.g., a CDN \cite{Florance16}. 

We are now interested in the expected overall yearly energy consumption of the online video service. We take the four typical online video services that are shown in Table~\ref{tab:energyExamplesOverview} as examples: an on-demand video service, a social network, an Internet protocol television (IPTV) service, and teleconferencing. More information on the choice of the exact values for energy consumption modeling is provided in detail in Appendix~\ref{app:energyModelParams}. 

The estimated overall yearly energy consumption of the four services is listed in Table~\ref{tab:energyExamplesOverview}. 
\begin{table}[t]
\caption{Examples of the overall yearly energy consumption of four different online video services.   }
\label{tab:energyExamplesOverview}
\vspace{-.4cm}
\begin{center}
\begin{tabular}{l||r||r|r|r}
\hline
System $s$ & Overall $E_s$ & $E_{s,\mathrm{UT}}$ & $E_{s,\mathrm{VP}}$ & $E_{s,\mathrm{NW}}$\\
\hline
On-demand      & $3.77\,$TWh & $3.65\,$TWh & $61.7\,$GWh & $60.2\,$GWh\\
IPTV           & $3.83\,$TWh & $3.65\,$TWh & $111\,$GWh  & $65.7\,$GWh\\
Social Network & $4.87\,$TWh & $94.6\,$GWh & $4.76\,$TWh & $14.1\,$GWh\\
Teleconference & $1.29\,$TWh & $1.22\,$TWh & $6.02\,$GWh & $56.9\,$GWh\\
\hline
\end{tabular}
\end{center}
\end{table}
We can see that the overall energy consumption $E_{s}$ is in the same order of magnitude for all considered services (between $1\,\mathrm{TWh}$ and $4\,\mathrm{TWh}$) and corresponds to roughly one percent of the total electrical power production of Germany in the year 2019 \cite{cleanenergywire20}, which corresponds to GHG emissions of approximately $2.5$ million tons $\mathrm{CO}_2\mathrm{E}$ \cite{cleanenergywire20_2}. Please note that these values highly depend on the chosen parametrization of the online video services. Hence, we cannot conclude that a certain service is more energy intensive than another.  

The overall energy consumption $E_s$ is then split up into the energy consumption caused by the end-user devices $E_{s,\mathrm{UT}}$, the DCs on the provider side $E_{s,\mathrm{VP}}$, and the transmission networks $E_{s,\mathrm{NW}}$ (cf. Table~\ref{tab:energyExamplesOverview}). We can see that the main energy consumption for on-demand video, IPTV, and teleconferencing services is caused by end-user devices, which account for $97\%$, $95\%$, and $95\%$ of the total energy consumption, respectively. For the social network, the energy consumption of the video providers is the most important component ($98\%$). It is worth mentioning that the energy consumption of the transmission networks is rather small for all services (below $100\,$GWh) and in each case corresponds to less than $5\%$ of the total energy consumption.  

\begin{figure}[t]
\centering
\psfrag{019}[bc][bc]{Yearly Energy Consumption}%
%
%
\psfrag{000}[ct][ct]{ \rotatebox{90}{Smartphones}}%
\psfrag{001}[ct][ct]{ \rotatebox{90}{Tablet PCs}}%
\psfrag{002}[ct][ct]{ \rotatebox{90}{Laptops}}%
\psfrag{003}[ct][ct]{ \rotatebox{90}{TVs}}%
\psfrag{004}[ct][ct]{ \rotatebox{90}{Desktop PCs}}%
\psfrag{005}[rc][rc]{ Wh}%
\psfrag{006}[rc][rc]{ kWh}%
\psfrag{007}[rc][rc]{ MWh}%
\psfrag{008}[rc][rc]{ GWh}%
\psfrag{009}[rc][rc]{ TWh}%
\psfrag{010}[rc][rc]{ }%
\psfrag{011}[bc][tc]{Yearly Energy Consumption }%
\psfrag{data1}[l][l]{\small On-demand}
\psfrag{data2}[l][l]{\small IPTV}
\psfrag{data3}[l][l]{\small Social Network}
\psfrag{dataaaaaaaaaaaaa1}[l][l]{\small Teleconference}
 \includegraphics[width=.38\textwidth]{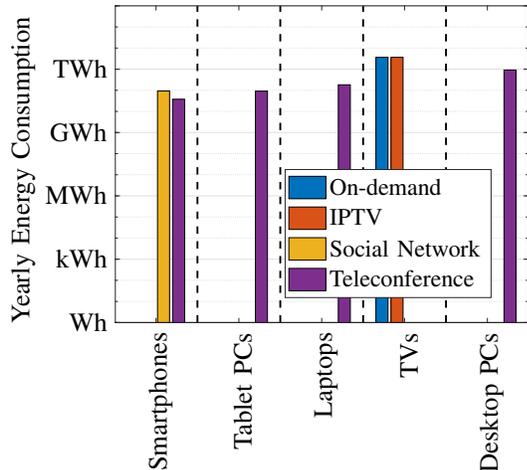} 
\vspace{1.cm}
\caption{Energy consumption from the end user's point of view depending on the service (color) and end-user device (horizontal axis). The vertical axis shows the yearly energy consumption in Wh and is plotted on a logarithmic scale. }
\label{fig:overview_UT}
\end{figure}

For a profound discussion and analysis, we further split up the energy consumption down to the device level. On the end-user side, we study the impact of the type of end-user device on  the energy consumption, where we consider smartphones, tablet PCs, laptops, TVs, and desktop PCs as listed in Tables~\ref{tab:smart} and \ref{tab:otherUTs}. Hence, using Eq. \eqref{eq:serviceWiseUTEnergy} we take subsets from the set of devices $D_s$, where each subset only contains one type of device. The resulting energy consumption values are illustrated in Fig.~\ref{fig:overview_UT}, where missing bars indicate that in our parametrization, the service is not requested from the corresponding device (cf. Appendix~\ref{app:energyModelParams}). 
We can see that most energy is consumed by TVs and desktop PCs, which is expected because their power consumption during operation is the highest. If devices with lower power consumption are used (e.g., smartphones or tablet PCs for a social network or teleconferencing), also the overall energy consumption is decreased by at least one order of magnitude. Hence, a significant amount of energy can be saved by simply switching from power-intensive desktop PCs or TVs to energy-efficient devices such as laptops and tablet PCs. 

\begin{figure}[t]
\centering
\psfrag{014}[bc][bc]{Yearly Energy Consumption}%
%
%
\psfrag{000}[ct][ct]{ \rotatebox{90}{Offset}}%
\psfrag{001}[ct][ct]{ \rotatebox{90}{Tx} }%
\psfrag{002}[ct][ct]{ \rotatebox{90}{Rx}}%
\psfrag{003}[ct][ct]{ \rotatebox{90}{Copies}}%
\psfrag{004}[ct][ct]{ \rotatebox{90}{Decoding}}%
\psfrag{005}[ct][ct]{ \rotatebox{90}{Encoding}}%
\psfrag{006}[ct][ct]{ \rotatebox{90}{Storage}}%
\psfrag{007}[rc][rc]{ Wh}%
\psfrag{008}[rc][rc]{ kWh}%
\psfrag{009}[rc][rc]{ MWh}%
\psfrag{010}[rc][rc]{ GWh}%
\psfrag{011}[rc][rc]{ TWh}%
\psfrag{012}[rc][rc]{ }%
\psfrag{013}[bc][bc]{Yearly Energy Consumption}%
\psfrag{data1}[l][l]{\small On-demand}
\psfrag{data2}[l][l]{\small IPTV}
\psfrag{data3}[l][l]{\small Social Network}
\psfrag{dataaaaaaaaaaaaa1}[l][l]{\small Teleconference}
 \includegraphics[width=.49\textwidth]{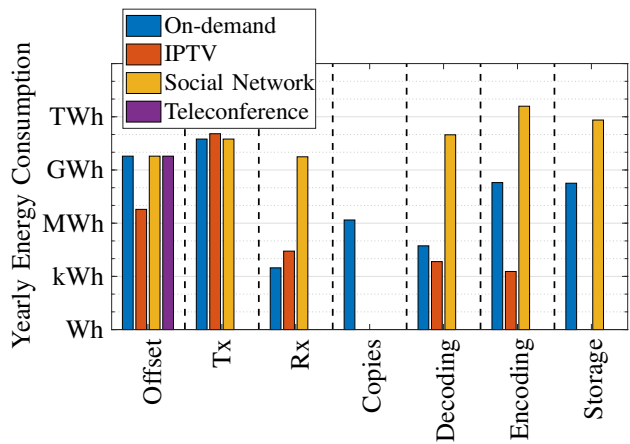} 
\vspace{.5cm}
\caption{Energy consumption from the video provider's point of view depending on service (color) and tasks (horizontal axis). The vertical axis shows the yearly energy consumption in Wh and is plotted on a logarithmic scale.  }
\label{fig:overview_VP}
\end{figure}

Considering the energy consumption on the provider side (VP), we split up the total energy consumption of the servers by focusing on the server tasks as shown in Eq. \eqref{eq:VP_serverParts}. Accordingly, Fig.~\ref{fig:overview_VP} illustrates the energy consumption corresponding to the offset, transmission and reception of videos, copy of videos in CDNs, decoding and encoding of videos, and storage of videos on surrogate servers. Missing bars indicate that the task is not performed by the corresponding service. 

We can see that for on-demand video streaming and IPTV, the energy required to send the videos (Tx) is the highest. For the social network, the energy consumption of encoding is crucial, which is also confirmed in the literature \cite{Wu20}. For the teleconferencing service, as most processing is performed at the end-user side, the main power consumption is related to the offset energy of the servers.

\subsection{Case Study for Energy Optimization}
\label{secsec:singleVideo}
Next to facilitating overall estimates, as shown above, the proposed  model allows the optimization of the overall energy efficiency on the video level. As an introductory example, we consider a system engineer working for a provider of freely accessible videos (tubes). His task is to choose optimal encoder settings for the lowest overall energy consumption. A new amateur movie called ``The Lord of the Streams'' is uploaded and shall be posted on the server. 
The video has a duration of two hours and is recorded with a high spatial and temporal resolution (4K at 60$\,$fps). 
The system engineer now needs to choose a suitable encoder. As we will see, the main difficulty of a good choice is that in advance, it is unknown how many end users will choose to watch the video, i.e., the number of devices $\left| D_s\right|$ requesting the video is unknown beforehand.  

When offering the video through the service, the system engineer now has to answer the following questions: 
\begin{itemize}
\item Which codec and bitrate to choose? 
\item How much energy to spend on encoding $(p_{s,\mathrm{VP},\sigma,\mathrm{trans},r,\mathrm{enc}}=\,?)$? 
\item Is it required to provide multiple streams at multiple bitrates coded in different codecs $(\left|V_{s,\mathrm{VP},\sigma,\mathrm{trans},r}\right|=\,?)$?
\item Is it required to store the video on surrogate servers and if so, on how many $(\left|\mathit{\Sigma_s}\right|=\,?)$?
\end{itemize}
For an illustration of the impact of these choices, we analyze the impact of two parameters: The encoding energy and the number of requests. From Table~\ref{tab:energyParamsVP}, we can choose two different encoding energy values. As indicated in \cite{Wu20}, the encoding energy has a significant impact on the bitrate  because the bitrate for high-energy encoding is reported to be $2.5$ times smaller than that for low-energy encoding. Hence, the two options are 
\begin{enumerate}
\item $p_{s,\mathrm{VP},\sigma,\mathrm{trans},r,\mathrm{enc}}=200\,\frac{\mathrm{mJ}}{\mathrm{s}_\mathrm{video}}$ and $B_r=80\,\mathrm{GByte}$,
\item  $p_{s,\mathrm{VP},\sigma,\mathrm{trans},r,\mathrm{enc}}=90\,\frac{\mathrm{kJ}}{\mathrm{s}_\mathrm{video}}$ and $B_r=32\,\mathrm{GByte}$.
\end{enumerate}

\begin{figure}[t]
\centering
\psfrag{019}[bc][bc]{Yearly Energy Consumption}%
%
%
\psfrag{000}[ct][ct]{$1$}%
\psfrag{001}[ct][ct]{}%
\psfrag{002}[ct][ct]{}%
\psfrag{003}[ct][ct]{$1{,}000$}%
\psfrag{004}[ct][ct]{}%
\psfrag{005}[ct][ct]{}%
\psfrag{006}[ct][ct]{$1\,$Mio}%
\psfrag{007}[ct][ct]{}%
\psfrag{008}[rc][rc]{$10\,$kWh}%
\psfrag{009}[rc][rc]{$1\,$MWh}%
\psfrag{010}[rc][rc]{$0.1\,$GWh}%
\psfrag{011}[tc][bc]{Number of Requests $\sum_{d\in D_s}\left|R_{s,\mathrm{UT},d}\right|$}%
\psfrag{012}[bc][tc]{Yearly Energy Consumption}%
\psfrag{dataaaaaaaaaaaaaaaaaaa1}[l][l]{\small High encoding energy}
\psfrag{data2}[l][l]{\small Low encoding energy}
 \includegraphics[width=.48\textwidth]{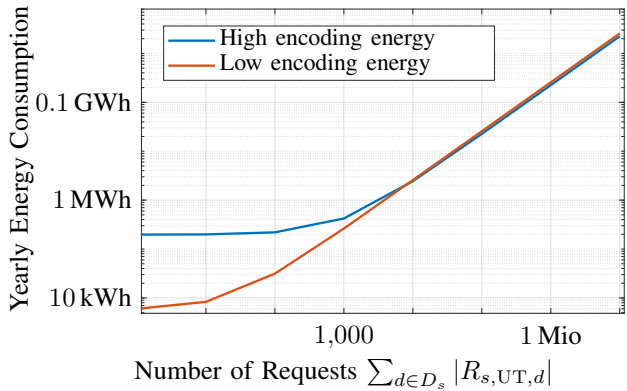} 
\caption{Yearly energy consumption (vertical axis) caused by a single video depending on the number of requests (horizontal axis). The blue line corresponds to high encoding energy and low bitrate, the red curve corresponds to low encoding energy and high bitrate.  }
\label{fig:numRequests}
\end{figure}

Using the same parametrization as for the on-demand video service in the last subsection, Fig.~\ref{fig:numRequests} shows the estimated overall energy consumption for this particular video depending on the number of requests $\sum_{d\in D_s}\left|R_{s,\mathrm{UT},d}\right|$. 

For a small number of requests (lower than $\sum_{d\in D_s}\left|R_{s,\mathrm{UT},d}\right| = 1{,}000$), it is apparently beneficial to choose the encoder with the low encoding energy consumption (red line) instead of the encoder with the high energy consumption (blue line), because the overall yearly energy consumption is much smaller. 
 However, for a large number of requests ($\sum_{d\in D_s}\left|R_{s,\mathrm{UT},d}\right| > 10{,}000$), the red line crosses the blue line such that energy-intensive encoding is preferable for a reduction of the overall energy consumption. The reason is that for this high number of requests, the energy savings due to the smaller number of bits to be transmitted to millions of users are higher than the encoding energy consumption.

To give an example, assuming  one million requests, $31.8\,$MWh can be saved with the energy-intensive encoder, which corresponds to GHG emissions of roughly $11.1\,$t$\,\mathrm{CO_2E}$ {\color[rgb]{0,0,0}(at $350\, \frac{\mathrm{gCO_2E}}{\mathrm{kWh}}$). 
On the other hand, if there is only a single request for a video, one can save $191\,$kWh by choosing the low energy encoder, which corresponds to GHG emissions of roughly $66.8\,$kg$\,\mathrm{CO_2E}$.

With this information, we can take this example one step further. Assume that the system engineer needs to handle $50{,}000$ videos, where $10$ videos are requested one million times and the remaining videos just once. We consider three cases: The system engineer can first encode all videos with the high-energy encoder, second with the low-energy encoder, and third, he can choose the energetically optimal encoder for all videos. The resulting overall energy consumption values and the corresponding GHG emissions are summarized in Table~\ref{tab:fullSystemOptimum}. 
\begin{table}[t]
\caption{Estimated overall energy consumption and GHG emissions for three different encoding configurations.   }
\label{tab:fullSystemOptimum}
\vspace{-.4cm}
\begin{center}
\begin{tabular}{l|r|r}
\hline
Encoder & Overall $E_s$ & GHG emissions\\
\hline
High energy      & $52.8\,$GWh & $18{,}480\,\mathrm{t\, CO_2E}$\\
Low energy      & $46.7\,$GWh & $16{,}345\,\mathrm{t\, CO_2E}$\\
Optimal      & $43.2\,$GWh & $15{,}120\,\mathrm{t\, CO_2E}$\\
\hline
\end{tabular}
\end{center}
\end{table}

The results show clearly that by optimally configuring the encoder, potential overall energy savings of almost $10\%$ can be achieved, if a low-energy encoder was used before. Comparing the optimized overall energy consumption with the case of high-energy encoding, the savings even exceed $20\%$. 

}

This simple example also shows that the proposed model allows for more sophisticated optimization. Many encoders provide multiple so-called presets, which allow choosing a tradeoff between the encoding energy consumption and the bitrate of a video \cite{Wu20,x265,x264}. With these presets, the system engineer can optimize the overall energy consumption depending on the expected number of requests. Furthermore, if the number of requests cannot be predicted accurately, time-adaptive solutions could be developed, where in the beginning, low-energy encoding is chosen and another energy-intensive encoding step is performed if end-user requests increase significantly.

{\color[rgb]{0,0,0}
In addition, we model the impact of providing multiple streams for a single video, i.e., we study the impact of $\left|V_{s,\mathrm{VP},\sigma,\mathrm{trans},r}\right|$ on the overall energy consumption. As system engineers, we can provide a set of bit streams encoded from a single video, where each bit stream is coded using different encoder settings. This might be required to satisfy the needs of different end users and their devices. The resulting set of bit streams can, e.g., differ by the bitrate, the codec, or the resolution. Let's assume that we encode sixteen bit streams that were created by encoding at two different bitrates, four different codecs, and two different resolutions. We further assume that these bit streams are stored on surrogate servers for availability to end users. We use the same parametrization as above using high-energy encoding and we assume to operate $10$ surrogate servers. We find that the overall energy consumption when providing a set of bit streams with respect to the provision of a single bit stream increases by a relatively small percentage of $0.07\%$ (assuming one million requests). However, this percentage changes drastically when the number of requests is lower. For example, if we only assume $1{,}000$ requests, the overall energy increases almost sixfold when providing $16$ streams instead of a single stream. This example underlines the importance of smart encoding decisions depending on the number of user requests. 

As a final example, if we scale the number of surrogate servers $\left|\mathit{\Sigma_s}\right|$, the energy consumption needed for maintenance of the CDN increases linearly. However, as the energy to store a stream and to distribute a stream to multiple servers is relatively small, the impact on the overall energy consumption is rather small (below $5\%$ even for a low number of requests). 
}

\begin{table*}[h!t!]
\caption{Parameters for online video energy consumption and their influence on the energy consumption of end-user devices, video providers' DCs, and transmission networks.  }
\label{tab:variables}
\vspace{-.4cm}
\begin{center}
\footnotesize
{\begin{tabular}{l|c|c|c|c|c|c|c|c}
\hline
  & \multicolumn{2}{c|}{End-user Terminals (UT)} & Networks & \multicolumn{5}{c}{ Video Providers (VP)} \\
Parameter 
 & Tx & Rx & (NW)  & Rx & transcode & copy  & store & Tx \\
\hline
 Hardware setup and configuration 
  & $\surd$ & $\surd$& $\surd$& $\surd$& $\surd$& $\surd$& $\surd$& $\surd$ \\
 Software setup and configuration 
  & $\surd$ & $\surd$& $\surd$& $\surd$& $\surd$& $\surd$& $\surd$& $\surd$ \\
 Network and configuration (wireless, wired, ...) 
 & $\surd$ & $\surd$& $\surd$& $\surd$& - &
  $\surd$& - & $\surd$\\
Number of uploaded videos $\sum_\sigma\left|R_{s,\mathrm{VP},\sigma,\mathrm{Rx}}\right|$ & $\surd$ & -& $\surd$& $\surd$& $\surd$& $\surd$ & $\surd$ & -\\
 Number of provided videos $\left| R_{s,\mathrm{VP},\sigma,\mathrm{trans}}\right| \cdot \sum_r\left| V_{s,\mathrm{VP},\sigma,\mathrm{trans},r}\right|$ & - & - & $\surd$ & - & $\surd$& $\surd$& $\surd$&- \\
 Number of streaming requests $\sum_d\left|R_{s,\mathrm{UT},d}\right|$ & - & $\surd$& $\surd$& - & - & - & - &$\surd$ \\
 Number of servers in CDN $\left|\mathit{\Sigma}_s\right|$ & - & - & $\surd$& - & -& $\surd$& $\surd$& - \\
Bitrate and file size of a requested video $b$, $B$  & $\surd$ & $\surd$& $\surd$& $\surd$& $\surd$& $\surd$& $\surd$& $\surd$ \\
 Duration of a video $t$  & $\surd$ & $\surd$& -& $\surd$& $\surd$& $\surd$& $\surd$ & $\surd$\\
 Other video properties (resolution,  frame rate, bit depth, codec)
 & $\surd$ & $\surd$& - & - & $\surd$& -& -& - \\
  Brightness of the end-user display  & - & $\surd$ & - &  - & - & - & - & - \\
 \hline
 \end{tabular}}
\end{center}
\end{table*}

\subsection{Directions for Future Research}
\label{secsec:approaches}

In the previous subsection, we have seen that the overall energy consumption of a single online video service can be optimized by choosing adequate encoder settings. In general, this idea can be applied to other tasks and systems involved in online video. To this end, the collected dependence of the energy consumption of online video systems on the main parameters used for modeling has to be identified. An overview of such dependencies is provided in Table~\ref{tab:variables}. The table shows which parameters influence the energy consumption of the considered end-user devices, the networks, and the server tasks on the provider side.

Further research targeting the energy efficiency of online video services is summarized in Table~\ref{tab:EnergyOptims}. 
\begin{table}[t]
\caption{Past research in the field of energy-efficient video streaming solutions.}
\label{tab:EnergyOptims}
\vspace{-.4cm}
\begin{center}
\footnotesize
{\begin{tabular}{l|c|r}
\hline
System & Attribute & References \\
\hline
Smartphone & UT & \cite{Herglotz20, Herglotz19b, Liu16,Zhang18}\\
Camera & UT & \cite{Saffari19, Lee12, Kleihorst01}\\
Decoder & UT & \cite{Herglotz19,Mallikarachchi20,Correa18,Li12}\\
Encoder & VP & \cite{Liang04,Kim11,He05,Penny16,Wu20}\\
CDN & VP & \cite{Chan11,Bianco16,Boscovic11,Goudarzi20}\\
 Mobile Networks & NW & \cite{Wu15,Miao10,Gyli11}\\
 \hline
 \end{tabular}}
\end{center}
\end{table}
The table shows that for end-user devices, networks, and servers, a considerable amount of research was performed to increase their energy efficiency. However, some topics are still to be addressed, which are discussed in the following. 

First, we can find that many studies do not represent the state of the art anymore because the investigated devices and systems are outdated and often not used anymore. Examples for such studies are \cite{Malmodin14,Herglotz13,Bianco16,Carroll13}. For future research, updates for these studies with {\color[rgb]{0,0,0} measurements on } state-of-the-art devices and systems would be highly interesting for the scientific community.  

Concerning the end users, more research on different types of end-user devices needs to be done. For example, to the best of the authors' knowledge, there are no detailed studies on the power consumption of TVs performing online video streaming. Similar to studies for smartphones or laptops \cite{Herglotz20,Carroll13,Herglotz22a}, such studies could produce more detailed information and opportunities to further increase the energy efficiency of TVs on the algorithmic level. 

Considering server tasks on the provider side, first, the energy efficiency of encoding could be analyzed in more detail. Sophisticated energy models similar to models available for the decoding energy \cite{Herglotz18} could be developed and applied to reduce the encoder's energy consumption. Second, from a high-level point of view, the energy consumption of servers including transmission, decoding, and storage could be addressed to increase the overall energy efficiency of an online video provider. 

Finally, a promising direction of research would be the optimization of an entire online video service as indicated in Subsection~\ref{secsec:singleVideo}. {\color[rgb]{0,0,0} For this, all three components (end-user devices, transmission networks, and provider-side servers) could be considered simultaneously and variable parameters such as the video properties, the number of provided videos, and the construction of the CDN could be optimized. In terms of video properties, parameters such as the bitrate, the target visual quality, the encoder implementation as well as configuration, the codecs, the resolution, and the frame rate could be considered. } These examples show the high potential and need for further research targeting energy efficiency in online video, which is  essential for sustainable deployment of online video technology in the future.

%
%
%
%
%
%
%
%
%
%
%
%
%
%

\section{Conclusion}
\label{sec:concl}
This paper has given an overview of the energy consumption of online video services, where links to greenhouse gas emissions causing climate change were established. In this regard, as a first contribution, we reviewed the current knowledge and recent research targeting the energy consumption analysis and energy modeling of online video systems. Subsequently, we constructed an energy model for online video services, which is a helpful design tool for understanding and optimizing the greenhouse gas emissions caused by a video streaming system. Finally, we discussed the implications of the developed model and provided directions for future research, which shall, e.g., target the holistic optimization of video streaming services by taking into account, among others, the expected number of users, the choice of the encoder, or the structure of a content delivery network.

\appendices

\section{Overview of Parameters and Sets for Energy Modeling}
This appendix provides an overview of all parameters employed in the proposed energy model. Table~\ref{tab:sets} lists all sets (end-user devices, servers, end-user requests, server requests, transcoding requests, and video sets) and Table~\ref{tab:params} lists all parameters that are used in this work (energy values, power values, server properties, and video properties). 
\begin{table*}
\caption{Sets for energy modeling.  }
\label{tab:sets}
\vspace{-.4cm}
\begin{center}
\begin{tabular}{l|c|l}
\hline
Set & Index & Meaning \\
\hline
$S$ & $s$ & Set of online video services (e.g., on-demand, IPTV, social network). \\
$D_s$ & $d$ & Set of end-user devices that send video requests to the providers (e.g., smartphones, tablet PCs, TVs).\\
$\mathit{\Sigma}_s$ & $\sigma$ & Set of provider-side servers providing online videos. \\
$R_{s,\mathrm{UT},d}$ & $r$ & Set of requests from the end user's device $d$. The request can include an upstream, a downstream, or both at the same time. \\
$R_{s,\mathrm{VP},\sigma,\mathrm{Tx}}$ & $r$ & Set of requests to stream a video to users, which is directed to the $\sigma$-th server. \\
$R_{s,\mathrm{VP},\sigma,\mathrm{copy}}$ & $r$ & Set of requests to copy a video to the $\sigma$-th surrogate server. \\
$R_{s,\mathrm{VP},\sigma,\mathrm{Rx}}$ & $r$ & Set of requests to receive a video, which is directed to the $\sigma$-th server. \\
$R_{s,\mathrm{VP},\sigma,\mathrm{trans}}$ & $r$ & Set of requests to transcode an incoming video, which is directed to the $\sigma$-th server. \\$V_{s,\mathrm{VP},\sigma,\mathrm{trans},r}$ & $v$ & Set of video versions to encode from the $r$-th incoming video, which is directed to the $\sigma$-th server. \\
$V_{s,\mathrm{VP},\sigma,\mathrm{store}}$ & $v$ & Set of videos stored on the $\sigma$-th server. \\
\hline
\end{tabular}
\end{center}
\end{table*}

\begin{table*}
\caption{Parameters for energy modeling. }
\label{tab:params}
\vspace{-.4cm}
\begin{center}
\begin{tabular}{l|c|l}
\hline
Parameter & Unit & Meaning \\
\hline
\multicolumn{3}{c}{\textbf{Global parameters}}\\
$\boldsymbol{E}$ & $\left[\frac{\mathrm{kWh}}{\mathrm{year}}\right]$ & Global yearly energy consumption of all online video services.\\
$E_s$ & $\left[\frac{\mathrm{kWh}}{\mathrm{year}}\right]$& Energy consumption caused by the $s$-th online video service. \\
$E_{s,\mathrm{UT}}$ & $\left[\frac{\mathrm{kWh}}{\mathrm{year}}\right]$& Total energy consumption of end-user devices using the $s$-th online video service. \\
$E_{s,\mathrm{VP}}$ & $\left[\frac{\mathrm{kWh}}{\mathrm{year}}\right]$& Energy consumption caused by the video providers' servers of the $s$-th online video service. \\
$E_{s,\mathrm{NW}}$ & $\left[\frac{\mathrm{kWh}}{\mathrm{year}}\right]$& Energy consumption of transmission networks caused by the $s$-th online video service. \\
\hline
\multicolumn{3}{c}{\textbf{Parameters on the end-user side}}\\
$E_{s,\mathrm{UT},d}$ & $\left[\frac{\mathrm{kWh}}{\mathrm{year}}\right]$& Energy consumption of the $d$-th end-user device using the $s$-th online video service. \\
$E_{s,\mathrm{UT},d,r}$& $\left[\mathrm{kWh}\right]$ & Energy consumption of the $d$-th end-user device caused by the $r$-th request. \\
$p_{s,\mathrm{UT},d,r,0}$& $\left[\mathrm{W}\right]$ & Offset power of end-user device $d$ for request $r$ (e.g., smartphone, tablet PC, TV). \\
$p_{s,\mathrm{UT},d,r,\mathrm{Rx}}$ & $\left[\mathrm{W}\right]$& Power of end-user device $d$ for request $r$, which is attributed to receiving a video.  \\
$p_{s,\mathrm{UT},d,r,\mathrm{Tx}}$ & $\left[\mathrm{W}\right]$& Power of end-user device $d$ for request $r$, which is attributed to the transmission (sending) of a video.\\
\hline

\multicolumn{3}{c}{\textbf{Parameters on the provider side}}\\
$\eta_\sigma$ & $\left[-\right]$& Power usage effectiveness (PUE) of the $\sigma$-th server. \\
$E_{s,\mathrm{VP},\sigma}$ & $\left[\frac{\mathrm{kWh}}{\mathrm{year}}\right]$& Energy consumption of the $\sigma$-th server of the $s$-th online video service. \\
$E_{s,\mathrm{VP},\sigma,0}$ & $\left[\frac{\mathrm{kWh}}{\mathrm{year}}\right]$& Offset energy consumption of the $\sigma$-th server.\\
$E_{s,\mathrm{VP},\sigma,\mathrm{Rx}}$ & $\left[\frac{\mathrm{kWh}}{\mathrm{year}}\right]$& Energy consumption of the $\sigma$-th server caused by receiving videos.\\
$E_{s,\mathrm{VP},\sigma,\mathrm{trans}}$ & $\left[\frac{\mathrm{kWh}}{\mathrm{year}}\right]$& Energy consumption of the $\sigma$-th server caused transcoding of videos.\\
$E_{s,\mathrm{VP},\sigma,\mathrm{copy}}$ & $\left[\frac{\mathrm{kWh}}{\mathrm{year}}\right]$& Energy consumption of the $\sigma$-th server caused by copying videos from a main server (only in CDNs).\\
$E_{s,\mathrm{VP},\sigma,\mathrm{store}}$ & $\left[\frac{\mathrm{kWh}}{\mathrm{year}}\right]$& Energy consumption of the $\sigma$-th server caused by storage of videos.\\
$E_{s,\mathrm{VP},\sigma,\mathrm{Tx}}$ & $\left[\frac{\mathrm{kWh}}{\mathrm{year}}\right]$& Energy consumption of the $\sigma$-th server caused by sending videos to end users (or to surrogate servers in CDNs).\\
$e_{s,\mathrm{VP},\sigma,\mathrm{send}}$ & $\left[\frac{\mathrm{kWh}}{\mathrm{bit}}\right]$& Energy consumption of the $\sigma$-th server for sending one bit. \\
$e_{s,\mathrm{VP},\sigma,\mathrm{Rx}}$ & $\left[\frac{\mathrm{kWh}}{\mathrm{bit}}\right]$ &Energy consumption of the $\sigma$-th server for receiving one bit.\\
$e_{s,\mathrm{VP},\sigma,\mathrm{trans},r,\mathrm{dec}}$ & $\left[\mathrm{kWh}\right]$ & Energy consumption of the $\sigma$-th server for decoding the video corresponding to the $r$-th transcoding request.\\
$p_{s,\mathrm{VP},\sigma,\mathrm{trans},r,\mathrm{dec}}$ & $\left[\mathrm{W}\right]$ & Decoding power consumption corresponding to the decoding energy consumption $e_{s,\mathrm{VP},\sigma,\mathrm{trans},r,\mathrm{dec}}$.\\
$e_{s,\mathrm{VP},\sigma,\mathrm{trans},r,\mathrm{enc},v}$ & $\left[\mathrm{kWh}\right]$ & Energy consumption of the $\sigma$-th server for encoding the $v$-th video from the $r$-th transcoding requests.\\
$p_{s,\mathrm{VP},\sigma,\mathrm{trans},r,\mathrm{enc},v}$ & $\left[\mathrm{W}\right]$ & Encoding power consumption corresponding to the encoding energy consumption $e_{s,\mathrm{VP},\sigma,\mathrm{trans},r,\mathrm{enc},v}$.\\
$e_{s,\mathrm{VP},\sigma,\mathrm{store}}$ & $\left[\frac{\mathrm{kWh}}{\mathrm{year}\cdot \mathrm{bit}}\right]$& Energy consumption of the $\sigma$-th server for storing one bit.\\
\hline

\multicolumn{3}{c}{\textbf{Parameters on the network side}}\\
$E_{s,\mathrm{NW},\mathrm{UT}}$ & $\left[\frac{\mathrm{kWh}}{\mathrm{year}}\right]$& Energy consumption of transmission networks caused by end-user requests using the $s$-th online video service. \\
$E_{s,\mathrm{NW},\mathrm{CDN}}$ & $\left[\frac{\mathrm{kWh}}{\mathrm{year}}\right]$& Energy consumption of transmission networks caused by provider requests (for CDNs). \\
$p_{s,\mathrm{NW},\mathrm{UT},d,r,0}$ & $\left[\mathrm{W}\right]$ & Offset power consumption of the network for the $r$-th request of the $d$-th end-user device.\\
$p_{s,\mathrm{NW},\mathrm{UT},d,r,\mathrm{rate}}$ & $\left[\frac{\mathrm{W}}{\mathrm{bit}}\right]$ & Bitrate-variable power consumption of the network for the $r$-th request of the $d$-th end-user device.\\
$p_{s,\mathrm{NW},\mathrm{CDN},\sigma,r,0}$ & $\left[\mathrm{W}\right]$ & Offset power consumption of the network for the $r$-th request from the $\sigma$-th server.\\
$p_{s,\mathrm{NW},\mathrm{CDN},\sigma,r,\mathrm{rate}}$ & $\left[\frac{\mathrm{W}}{\mathrm{bit}}\right]$& Bitrate-variable power consumption of the network for the $r$-th request from the $\sigma$-th server.\\
\hline
\multicolumn{3}{c}{\textbf{Video parameters}}\\
$b$ & $\left[\frac{\mathrm{kbit}}{\mathrm{s}}\right]$& Bitrate of a video. \\
$B$ & $\left[\mathrm{bit}\right]$&  Size of a video. \\
$t$ & $\left[\mathrm{s}\right]$&Duration of a video.\\
\hline
\end{tabular}
\end{center}
\end{table*}

\section{Parameter Values for Energy Modeling}
\label{app:energyModelParams}
This appendix shows and explains the assumed parameter values for the estimations of the overall energy consumption of online video services in Subsection~\ref{secsec:serviceEnergy}. For all services, we assume that there are $100\,$ million users which employ $\left|D_{s}\right|=100\,$million devices. Each user makes use of the service one hour per day on average (cf. Subsection~\ref{secsec:serviceEnergy}). For the on-demand video service, we assume that all videos have a duration of two hours, i.e., $t_r = 7200\,$s. For the targeted average use of an hour a day, this results in one request every second day such that the number of requests per device is the size of the set of requests $\left|R_{s,\mathrm{UT},d}\right| = \frac{365}{2}$. Furthermore, we assume that all requests are performed from TV sets such that we take the power value for TVs from Table~\ref{tab:otherUTs}. We assume that the service provides $\left|R_{s,\mathrm{VP},\sigma,\mathrm{trans}}\right|=1000$ different videos to be requested by end users. These were encoded from $\left|R_{s,\mathrm{VP},\sigma,\mathrm{Rx}}\right|=1000$ input videos, which means that here, we adopt the simplified assumption that only a single version of each input video is provided 
($\left|V_{s,\mathrm{VP},\sigma,\mathrm{trans},r}\right|=1$). Furthermore, half of these videos are stored on $\mathit{\Sigma}_s-1=999$ surrogate servers of the CDN because they are frequently requested \cite{Bianco16} such that $\left|V_{s,\mathrm{VP},\sigma,\mathrm{store}}\right|=500$. For the network connection, as TV sets are used, we take transmission energy consumption values for fixed BB access networks and we assume that complex software transcoding is employed. 

For IPTV, we make similar assumptions. The main differences to on-demand services are that we assume a higher bitrate ($b_r=10\,\frac{\mathrm{Mbit}}{\mathrm{s}}$) because due to real-time restrictions, encoding can only be done in real time such that also the corresponding lower encoding power consumption from Table~\ref{tab:energyParamsVP} is employed. Furthermore, we assume that IPTV is constantly broadcasting content such that encoding of videos is performed $24$ hours on $365$ days of the year. Furthermore, we assume that the video is not stored.

For the social network service, we assume that all videos are recorded by end users. We assume that each uploaded video has a duration of $5$ minutes and that each video is watched by $10$ friends. When requesting one hour of video in total per day (which means that twelve videos are requested), the number of requests for downloading videos per device and per year is then $\left|R^\mathrm{down}_{s,\mathrm{UT},d}\right| =  365\cdot 12 = 4380$, where we assume that all requests are performed from smartphones. As each video is watched ten times, the corresponding number of uploaded videos is one-tenth of this number given by $\left|R^\mathrm{up}_{s,\mathrm{UT},d}\right| =  365\cdot 12 / 10 = 438$, such that on average, each user uploads more than one video per day. Furthermore, all uploaded videos are transcoded to a standard codec and container format such that $\left|R_{s,\mathrm{VP},\sigma,\mathrm{trans}}\right|=\left|D_s\right|\cdot \left|R^\mathrm{up}_{s,\mathrm{UT},d}\right|  = 43.8\cdot 10^9$. However, as all videos are only requested a few times, they are only stored on a single server (no copies are stored on surrogate servers). 
Still, a network of $\left|\mathit{\Sigma}_s\right|=1000$ servers ensures that all videos are stored close to the end users. Encoding and decoding are done using software implementations resulting in energy-intensive processing (cf. Table~\ref{tab:energyParamsVP}) at a bitrate of $b_r=5\,\frac{\mathrm{Mbit}}{\mathrm{s}}$. To take into account that social network providers will probably not use the most sophisticated and energy-intensive encoding process ($90\,\frac{\mathrm{kJ}}{\mathrm{s_{video}}}$), we choose a significantly lower value of $1\,\frac{\mathrm{kJ}}{\mathrm{s_{video}}}$. The network is chosen to be a mobile network.

For teleconferencing, we assume that end users use different devices to participate in conferences. For simplicity, we assume that the share of devices is evenly distributed among smartphones, tablet PCs, Laptops, and desktop PCs. Each end user is logged in one hour per day in a single conference such that $\left|R_{s,\mathrm{UT},d}\right| =  365$ and $t_r = 3600\,\mathrm{s}$. Furthermore, as high visual qualities are not as crucial for teleconferencing as for on-demand streaming, we assume that the resolution of the videos is significantly lower than HD, which results in a reduced bitrate of $2\,\frac{\mathrm{Mbit}}{\mathrm{s}}$. 

Finally, we consider the energy consumption of the transmission network using values from Table~\ref{tab:energyParamsNW}. We assume that the social network is accessed using a mobile BB access link and the other services use fixed BB access. The copying of videos via the CDN is taken into account using the values from the fixed BB access link. 

\section*{Acknowledgment}
We thank Wolfgang Heyn for his profound literature research, where he collected methods and data to describe the global energy consumption of online video services.

\ifCLASSOPTIONcaptionsoff
  \newpage
\fi

\bibliographystyle{IEEEtran}
\end{document}